\documentclass[aps,prb,reprint,unsortedaddress,showpacs,floatfix]{revtex4-1}
\usepackage{graphicx}
\usepackage{dcolumn}
\usepackage{bm}
\usepackage{subfigure}
\usepackage{amsmath}
\usepackage{amssymb}
\usepackage{nicefrac}
\begin{document}
\title{Comparing charge transfer tuning effects by chemical
  substitution and uniaxial pressure in the organic charge transfer
  complex tetramethoxypyrene-tetracyanoquinodimethane}
\author{Milan Rudloff} \author{Kai Ackermann} \author{Michael Huth}
\email{michael.huth@physik.uni-frankfurt.de}
\affiliation{Physikalisches Institut, Goethe-Universit\"{a}t,
  Max-von-Laue-Strasse 1, 60438 Frankfurt am Main, Germany}
\author{Harald O. Jeschke} \author{Milan Tomic} \author{Roser Valenti}
\affiliation{Institut f\"{u}r Theoretische Physik,
  Goethe-Universit\"{a}t, Max-von-Laue-Strasse 1, 60438 Frankfurt am
  Main, Germany}
\author{Benedikt Wolfram} \author{Martin Br\"{o}ring}
\affiliation{Institut f\"{u}r Anorganische und Analytische Chemie,
  Technische Universit\"{a}t Carolo Wilhelmina, Hagenring 30, 38106
  Braunschweig, Germany}
\author{Michael Bolte} \affiliation{Institut f\"{u}r Anorganische und
  Analytische Chemie, Goethe-Universit\"{a}t, Max-von-Laue-Strasse
  7,\\60438 Frankfurt am Main, Germany}
\author{Dennis Chercka}
\author{Martin Baumgarten}
\author{Klaus M\"{u}llen}
\affiliation{Max-Planck-Institut f\"{u}r Polymerforschung, Ackermannweg 10, 55021 Mainz, Germany}
\date{\today}
\begin{abstract}
  In the search for novel organic charge transfer salts with variable
  charge transfer degree we study the effects of two modifications to
  the recently synthesized donor-acceptor [Tetramethoxypyrene
  (TMP)]-[Tetracyanoquinodimethane (TCNQ)]. One is of chemical nature
  by substituting the acceptor TCNQ molecules by F$_4$TCNQ
  molecules. The second consists in simulating the application of
  uniaxial pressure along the stacking axis of the system. In order to
  test the chemical substitution, we have grown single crystals of
  TMP-F$_4$TCNQ and analyzed its electronic structure via electronic
  transport measurements, {\it ab initio} density functional theory
  (DFT) calculations and UV/VIS/IR absorption spectroscopy. This
  system shows an almost ideal geometrical overlap of nearly planar
  molecules alternately stacked (mixed stack) and this arrangement
  is echoed by a semiconductor-like transport behavior with an
  increased conductivity along the stacking direction. This is in
  contrast to TMP-TCNQ which shows a less pronounced anisotropy and a
  smaller conductivity response. Our bandstructure calculations
  confirm the one-dimensional behavior of TMP-F$_4$TCNQ with
  pronounced dispersion only along the stacking axis. Infrared
  measurements illustrating the C$\equiv$N vibration frequency shift
  in F$_4$TCNQ suggest however no improvement on the degree of charge
  transfer in TMP-F$_4$TCNQ with respect to TMP-TCNQ. In both
  complexes about 0.1 is transferred from TMP to the
  acceptor. Concerning the pressure effect, our DFT calculations on
  designed TMP-TCNQ and TMP-F$_4$TCNQ structures under different
  pressure conditions show that application of uniaxial pressure along
  the stacking axis of TMP-TCNQ may be the route to follow in order to
  obtain a much more pronounced charge transfer.
\end{abstract}
%
\maketitle
\section{Introduction}
Organic charge transfer (CT) systems offer both a playground for
studying fundamental solid state properties as well as  a material
class that in recent years has become extremely attractive for modern
organic electronic devices, such as organic thin film
transistors. \cite{applic1, applic2, applic3, applic4, applic5,
  applic6} The electrostatic bonding of two different organic
molecules, donor (D) and acceptor (A), gives rise to a rich electronic
behavior. This feature combined with the vast flexibility  chemists
have concerning molecular design makes it possible to gain access to
such phenomena as unconventional superconductivity, ferroelectricity,
spin liquid  or metal-to-insulator transitions, to mention a
few.\cite{1Dmetals,organConduct,organFerroelec} Charge transfer
compounds tend to form one- or two-dimensional structures (stacks or
layers) that lead to a simple differentiation between systems composed
of segregated stacks of the donor and acceptor and systems where donor
and acceptor are mixed within the stacks. The vast majority of CT
compounds are of the latter mixed-stack type and have long been
considered as a material class of reduced significance in basic and
applied research, as the geometrical arrangement and resulting
molecular orbital overlap leads to semiconducting or insulating
behavior. However, this view needs revision in several
respects. Ambipolar charge-transport properties have been predicted by
electronic structure calculations in selected mixed-stack systems with
mobility values that would rival the best single-component organic
semiconductors, such as rubrene or
pentacene.\cite{ambChargeTrans_mixedStack_JACS2012,bandstruktur_tetracen+perylenTCNQ}
Moreover, a growing group of mixed-stack CT systems shows a
temperature- and/or pressure driven transition into a ferroelectric or
an antiferroelectric state associated with a pronounced increase of
the charge transfer degree, a so-called neutral-ionic (NI)
transition.\cite{NIchemphys2006} The NI-transition is widely tunable
by pressure (in single crystals)\cite{Lemee1997a} or biaxial strain
(in clamped thin films)\cite{Huth2014a,vita_synthMet2011,ETTCNQ_vita}
and may even be associated with a multiferroic state in which a,
typically, antiferromagnetic order parameter couples to the electrical
polarization.\cite{Filatov2013a} With a view on strengthening the
material basis for this class of mixed-stack systems, several
challenges have to be overcome. Despite of the fact that most of the
organic CT systems crystallize in a mixed-stack variant, the
conditions for the establishment of a temperature-driven NI-transition
are rather stringent. Roughly speaking, starting from a moderate
initial charge transfer degree of about 0.3 to 0.4 from donor to
acceptor the gain of Madelung energy under thermal contraction as the
materials cools down must be sufficient to overcompensate the
generally still positive difference in the ionization energy of the
donor and the electronic affinity of the acceptor. Under these
conditions a spontaneous increase of the charge transfer degree,
usually in conjunction with D-A dimerization and a loss of inversion
symmetry, can occur leading to an (anti-) ferroelectric state. At the
beginning of a systematic identification of possible candidates for a
NI-transition lies the proper selection of the donor and acceptor
species based on their respective molecular energies, as has been
suggested by Torrance (V-shaped diagram).\cite{NI_discovery} It has to
be noted, though, that a good match of the molecular energies does not
guarantee a NI-transition to occur. By way of proper modification of
the functional groups of either donor or acceptor a fine-tuning of the
respective ionization energies and electron affinities can be
done. However, in co-crystallizing donor and acceptor a possible
change of the resulting crystal structure as compared to the mother
compound is hard to prevent and not predictable. Pressure-based tuning
of the mother compound, ideally leaving the crystal structure
unchanged, is an attractive alternative, albeit technically
non-trivial and thus not suitable as a screening method for the
identification of new NI-transition systems. Here the question arises
which approaches with anticipatory quality can be developed that allow
to identify possible candidates beforehand, once a promising mother
compound is identified.

In this work we present a combined experimental and theoretical study
on the electronic properties of the new mixed-stack organic CT
compound [Tetramethoxypyrene
(TMP)]-[Tetrafluoro-tetracyanoquinodimethane (F$_4$TCNQ)] which was
derived from the previously synthesized TMP-TCNQ mixed-stack system by
way of enhancing the acceptor qualities of TCNQ by halogen
substitution. The pyrene TMP and the TCNQ structure both offer a
multitude of possible functional modifications that allow to
investigate how specific changes affect the structural and electronic
properties of the CT compound.
\begin{figure}[htp]
\includegraphics[width=0.4\textwidth]{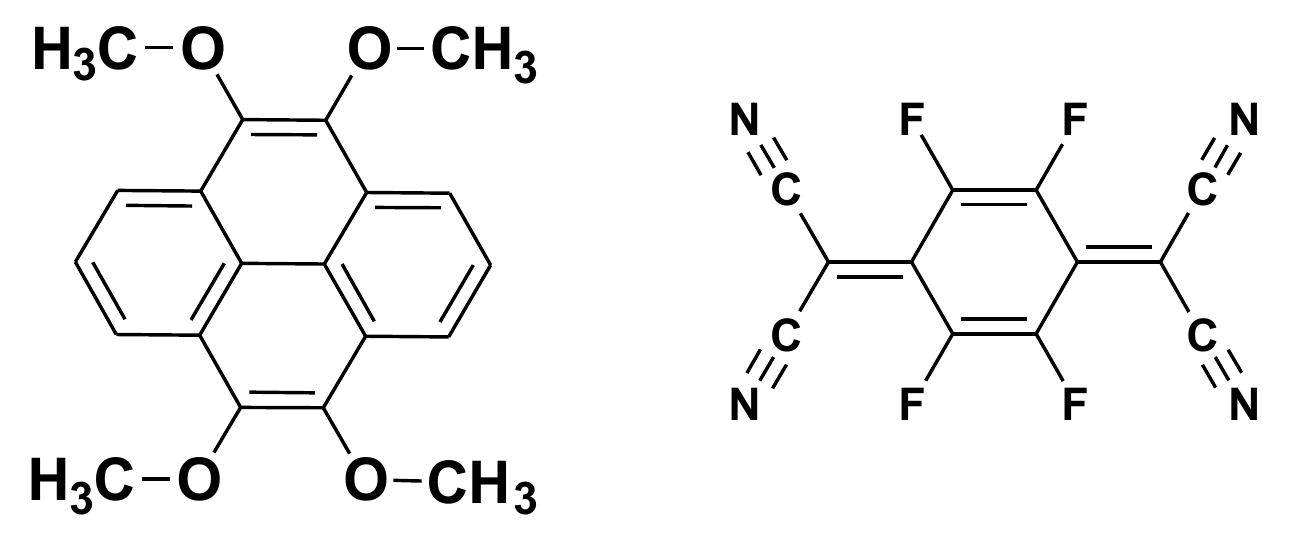}
\caption{Molecular structure of TMP (left) and F$_4$TCNQ (right). The
  four methoxy groups provide TMP with donor properties while the four
  cyano groups combined with four fluorine atoms give F$_4$TCNQ its
  acceptor functionality.}\label{fig:MolStruct}
\end{figure}
TMP-F$_44$TCNQ has a triclinic crystal structure with virtually ideal
geometrical overlap of donor and acceptor resulting in a pronounced
anisotropy of the electronic properties, as deduced from
temperature-dependent conductivity measurement and density functional
theory calculations. This structure differs from that of the mother
compound TMP-TCNQ which is far less anisotropic. Somewhat
counter-intuitively, the degree of charge transfer of TMP-F$_4$TCNQ is
not enhanced as compared to TMP-TCNQ but even reduced from about 0.2
to 0.1, according to the observed shifts in the C$\equiv$N stretching
mode as seen in IR absorption spectroscopy as well as in the
density functional theory (DFT) calculations. By analyzing the charge distribution
of the two complexes, we propose a possible explanation of these observations.
Finally by simulating the effect
of uniaxial pressure on these two compounds within the DFT
 calculations,  we predict a pronounced increase of the charge
transfer degree for TMP-TCNQ whereas TMP-F$_4$TCNQ appears to be
mostly unaffected. This band-structure-based approach of a simulated
pressure experiment may represent an efficient strategy to identify
possible candidates for NI-transition systems based on newly
synthesized mixed-stack organic charge transfer systems of suitable
initial charge transfer degree.
\section{Methods}
\label{sec:methods}
\subsection{Experimental methods}
\subsubsection{Synthesis and structure determination}
TMP was synthesized at the Max Planck Institute for Polymer Research
with a purity of better than 95\,$\%$. F$_4$TCNQ was purchased from
Sigma-Aldrich with a purity of $\geq$ 97\,$\%$.

Single crystals of TMP-F$_4$TCNQ were grown in solution by slow
evaporation of the solvent. TMP and F$_4$TCNQ were separately
solubilized in dichloromethane (DCM) in equimolar amounts (ca.\ 10
mg). Both solutions were carefully mixed in one beaker glass which was
then covered with aluminum foil to limit the evaporation rate. Black,
needle-like crystals could be extracted after one to two weeks.
 
The structure determination was based on a $0.41\times 0.06\times
0.06$\,mm sized crystal. X-ray data were recorded at T = 173\,K on a
STOE IPDS II two-circle-diffractometer using Mo-K$_\alpha$ radiation
and a Genix 3D multilayer optics monochromator. SHELXS97
\cite{SHELX_Sheldrick2008} was used for structure solution while
refinement was carried out with SHELXL97. Further information on data
collection and structure refinement are given in table
\ref{tab:xrayMeas_paramet}. Crystallographic data for TMP-F$_4$TCNQ
has been deposited with the Cambridge Crystallographic Data
Centre~\cite{ccdc}.

\begin{table}[ht]
\caption{\label{tab:xrayMeas_paramet} Measurement and refinement parameters.}
\begin{ruledtabular}
\begin{tabular}{ll}
measured reflections & 5551\\
$\theta$-range (deg) & 3.41-25.03\\
reflections used & 2305\\
parameters & 201\\
goodness of fit & 0.954
\end{tabular}
\end{ruledtabular}
\end{table}
\subsubsection{Electrical conductivity measurements}
Single crystals were placed onto a custom-made carrier chip and
electrically contacted with conducting graphite paste. 50\,$\mu$m
thick goldwires were attached to the paste and soldered to contact
pins on the carrier chip. On the small crystals two-probe measurements
proved to be suffficient, because the sample resistance dominated by
far the contact and wiring resistances. Low temperature measurements
were performed in an Oxford $^{4}$He cryostat with a Keithley 2636A
two-channel sourcemeter. Within the cryostat the carrier chip was
mounted in a variable temperature insert (VTI). The temperature was
measured with a calibrated resistance thermometer (Cernox).
\subsubsection{Infrared spectroscopy} 
Infrared spectroscopy was applied to powder material from solution
growth in the case of TMP-F$_4$TCNQ and the as-supplied powder in the
case of F$_4$TCNQ as a reference. Spectra were recorded at room
temperature with a Bruker Vertek 70 FT-IR spectrometer 
between 600 and
4000\,cm$^{-1}$. 
All measurements were performed with a resolution of 2\,cm$^{-1}$.
\subsubsection{UV-VIS/IR spectroscopy}
Spectra were recorded in solution (DCM) on a Perkin-Elmer Lambda 900
UV/VIS/NIR spectrometer. A concentration of 10$^{-4}$\,mol/l was used
for absorption between 0.1 and 1 at the wavelength region of
experimental interest.
\subsection{Theoretical methods}
We performed {\it ab initio} density functional theory calculations
for the TMP-F$_4$TCNQ complex. The full potential local orbital (FPLO)
basis set \cite{fplo} with the generalized gradient approximation
functional in its PBE form \cite{Perdew96} was used.  We employed a
$10\times 10\times 10$ $k$ mesh and analyzed the band structure by
fitting the bands near the Fermi level to a tight binding Hamiltonian
\cite{Kandpal2009}
\begin{equation}
H=\sum_i \varepsilon_i c_{i}^{\dag}c_{i} +\sum_{ij} t_{ij}c_{i}^{\dag}c_{j}
\label{eq:hamiltonian}
\end{equation}
with the operators $c_{i}^{\dag}$ ($c_{i}$) creating (annihilating)
electrons at site $i$, onsite energies $\varepsilon_i$ and transfer
integrals $t_{ij}$. We checked the uniqueness of the resulting fit
using projective Wannier functions.\cite{Jacko2013}

We simulate uniaxial pressure along the stacking direction by scaling
the $a$ lattice parameter of both TMP-TCNQ and TMP-F$_4$TCNQ in small
steps and fully relaxing all internal coordinates using the projector
augmented wave (PAW) basis and generalized gradient approximation
functional as implemented in the Vienna {\it ab initio} simulation
program (VASP).\cite{vasp1,vasp2} We use a $6\times 6\times 6$ $k$
mesh for TMP-TCNQ and a $4\times 4\times 4$ $k$ mesh for
TMP-F$_4$TCNQ, with a plane wave cutoff of 500\,eV.
\section{Results}
\subsection{Crystal structure}
\begin{figure}[htp]
\includegraphics[width=0.49\textwidth]{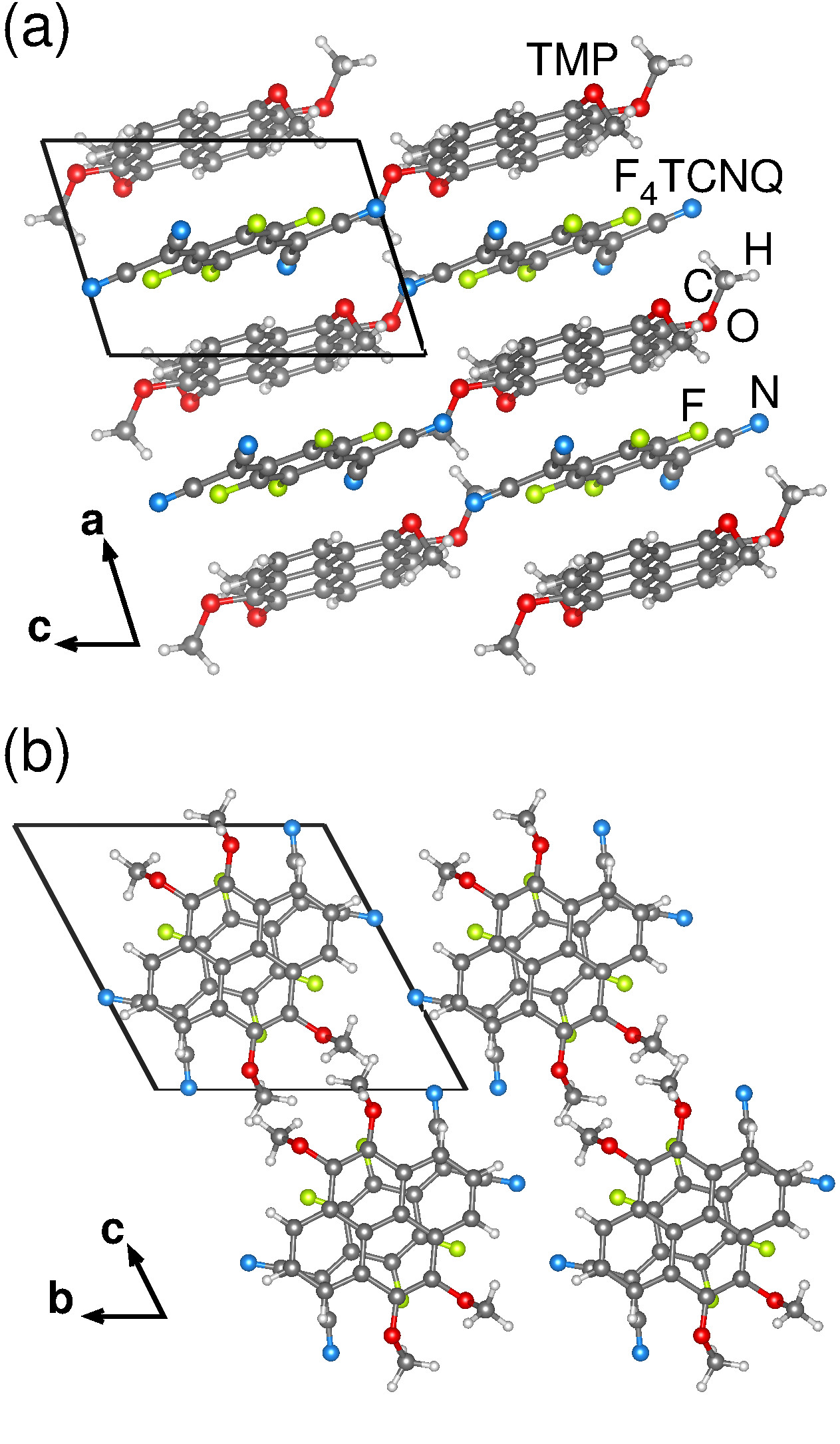}
\caption{Structure of TMP-F$_4$TCNQ viewed along (a) {\bf b}- and (b)
  {\bf a}-axis.}\label{fig:structures}
\end{figure}

The synthesized TMP-F$_4$TCNQ crystals showed a triclinic structure
with mixed stacks formed along its a-axis. Figure \ref{fig:structures}
illustrates the packing arrangement from different perspectives.
\begin{table}[ht]
\caption{\label{tab:crystalData} Crystal structure data for TMP-F$_4$TCNQ.}
\begin{ruledtabular}
\begin{tabular}{ll}
formula & C$_{20}$H$_{18}$O$_4$ - C$_{12}$F$_4$N$_4$\\
space group & $\mathrm{P}\overline{1}$\\
a [\AA{}] & 6.8078(13)\\
b [\AA{}] & 10.6592(18)\\
c [\AA{}] & 10.840(2)\\
$\alpha$ [$^\circ$] & 61.373(13)\\
$\beta$ [$^\circ$] & 71.958(14)\\
$\gamma$ [$^\circ$] & 84.063(14)\\
volume [\AA{}$^3$] & 655.6(2)\\
Z & 1
\end{tabular}
\end{ruledtabular}
\end{table}

All stacks have a common stacking axis. Along the stacks, TMP and
F$_4$TCNQ lie almost perfectly on top of each other: Their respective
planes (ignoring the methoxy groups) lie nearly perpendicular to the
stacking direction while the geometrical centers of TMP and F $_4$TCNQ
molecules are positioned on a common axis. The mean plane-to-plane
separation is about 3.35\,\AA{}.
\subsection{Electrical conductivity}
\begin{figure}[htp]
\includegraphics[width=0.43\textwidth]{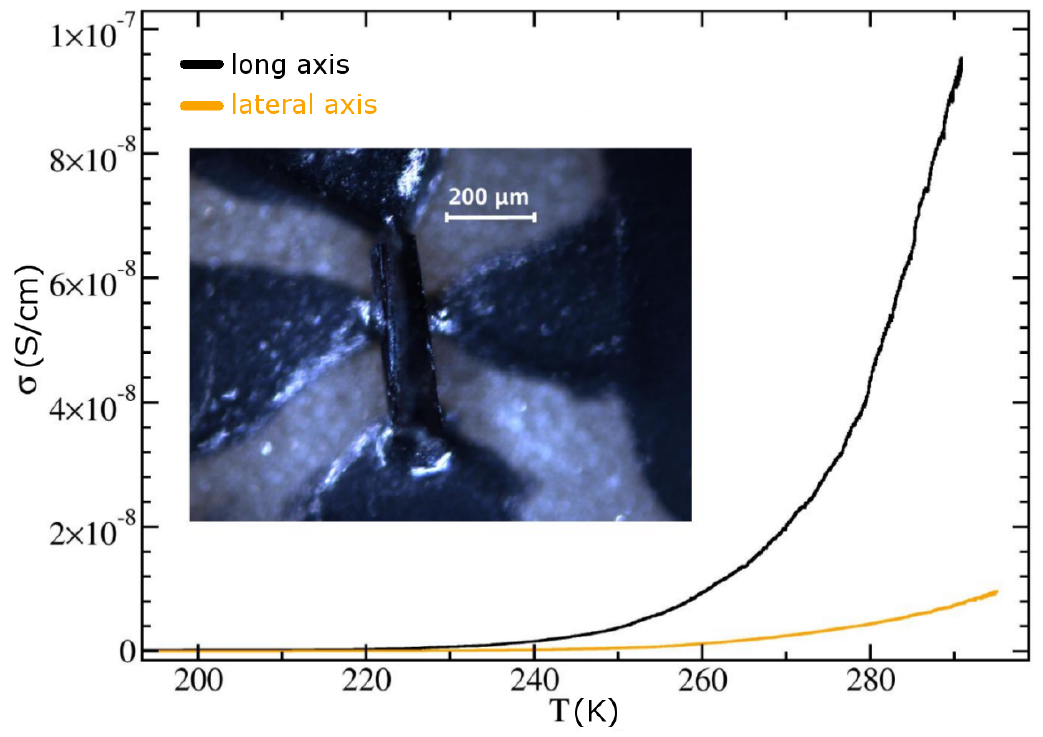}
\caption{Temperature-dependent conductivity measured on the
  TMP-F$_4$TCNQ crystal (shown with contacts in the inserted optical
  micrograph) for two different
  directions.}\label{fig:sigma_vs_T_+crystal}
\end{figure}

Electrical conductivity measurements for one typical crystal are
illustrated in Fig.~\ref{fig:sigma_vs_T_+crystal}. The crystal shown
here had dimensions of roughly $360\times 90\times
50$\,$\mu$m. Current-temperature curves were recorded with an applied
voltage of 100\,V, corresponding to a maximum electrical field of
11\,kV/cm, for two distinct directions along the crystal, one parallel
to its long axis and one perpendicular to it. The long axis
corresponded to the stacking direction of the molecules (a-axis).
\begin{figure}[htp]
\includegraphics[width=0.43\textwidth]{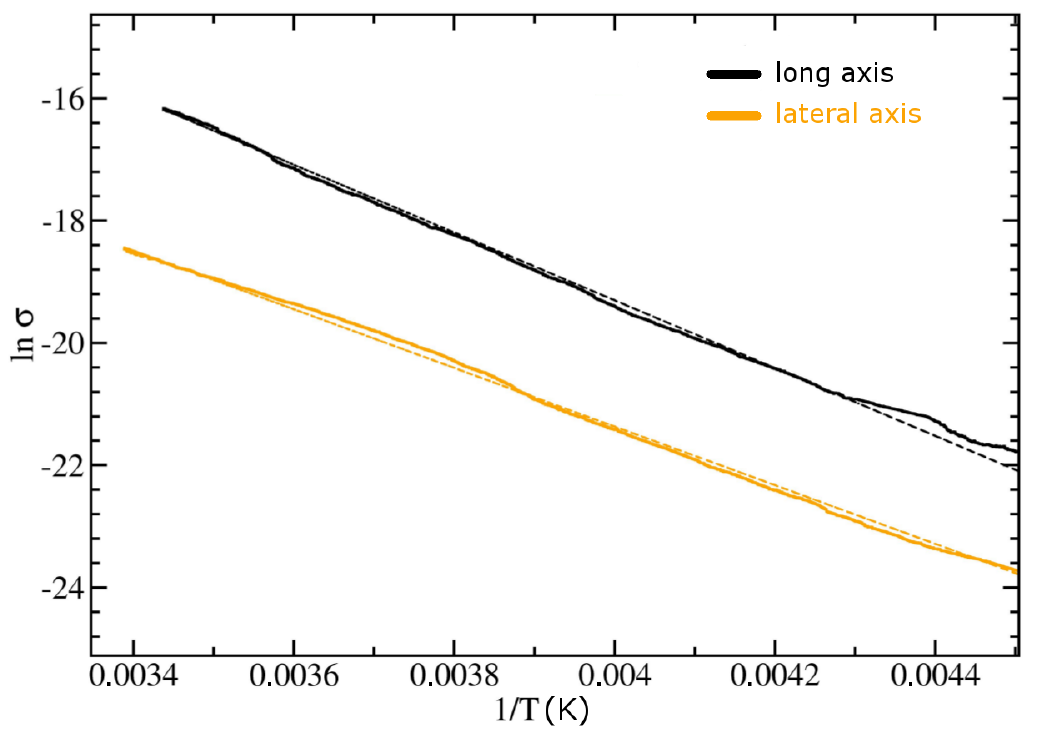}
\caption{Arrhenius plot and the corresponding linear fit (dashed
  lines) for both $\sigma$(T) measurements.}\label{fig:arrhenius}
\end{figure}

Both $\sigma$(T) characteristics show a thermally activated
(Arrhenius) behavior but with different activation energies E$_a$ (see
Fig.~\ref{fig:arrhenius}). This behavior was only visible at $T
\gtrsim 220$\,K as the resistance at lower temperatures became too
high for the measurement setup. For the stacking axis a value of E$_a
= 0.98$\,eV was calculated based on the fit performed in
Fig. \ref{fig:arrhenius}. The perpendicular direction gave a value of
E$_a = 0.84$ eV.
\subsection{Degree of charge transfer}
\begin{figure}[htp]
\includegraphics[width=0.43\textwidth]{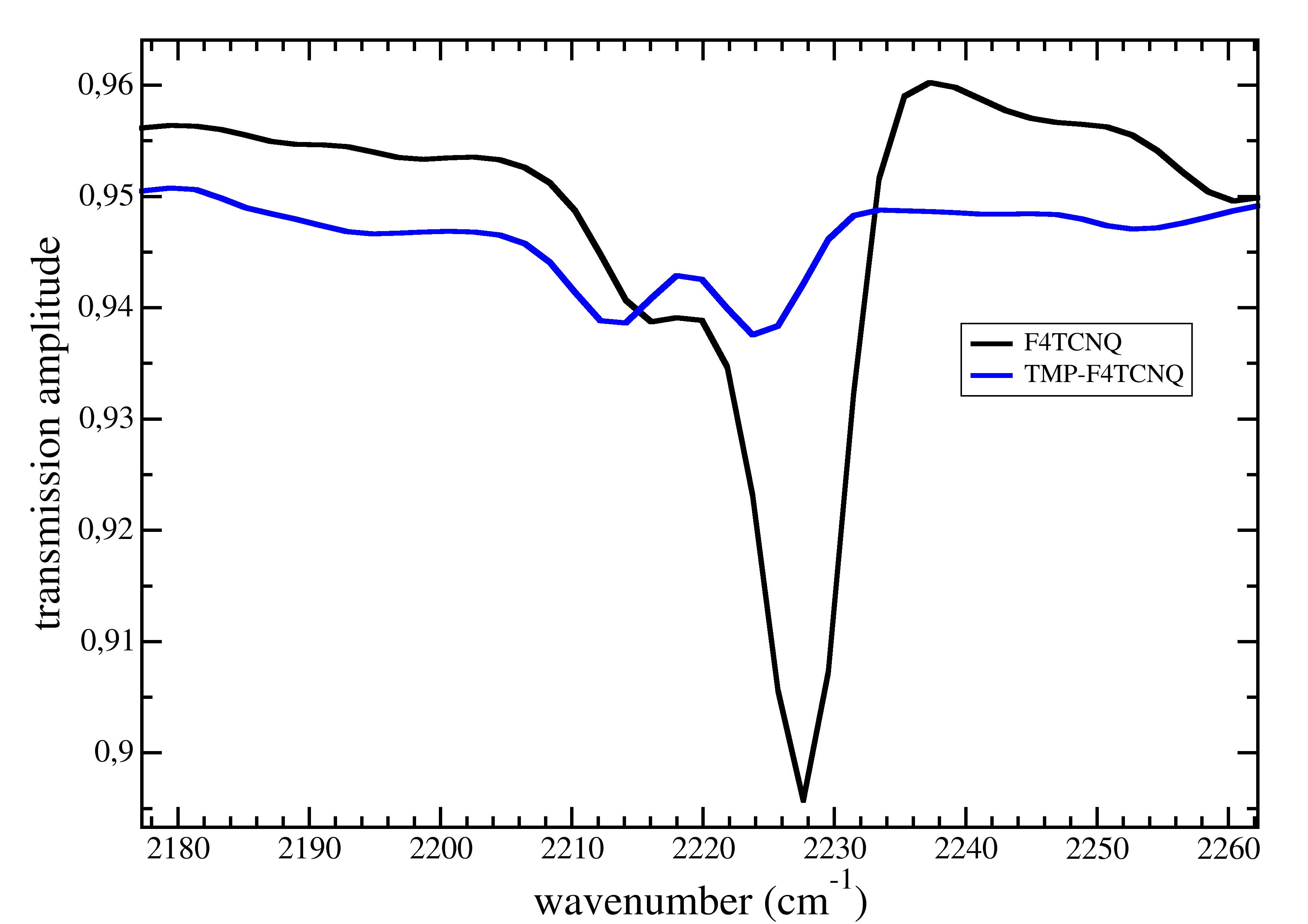}
\caption{IR absorption caused by the CN stretching vibrations in
  TMP-F$_4$TCNQ compared to the ones in F$_4$TCNQ. Both curves are an
  average over 16 scans.}\label{fig:IR_tmp+f4tcnq}
\end{figure}

The fraction of charge ($\rho$) transferred from donor to
acceptor (in the ground state) can be estimated by the change of the
C$\equiv$N stretching vibration frequency $\nu$ caused by the CT
complex formation.\cite{salmeronValverde1994,chappell1981} The
absorption of IR (infrared) radiation of isolated F$_4$TCNQ was
measured as a reference. The spectrum shown in
Fig. \ref{fig:IR_tmp+f4tcnq} illustrates IR bands at 2217 and
2228\,cm$^{-1}$ corresponding to the modes b$_{2u}$ and b$_{1u}$.

Both values are in close agreement with results from other
measurements.\cite{F4TCNQ_meneghetti,F4TCNQ_pingel} The spectrum for
TMP-F$_4$TCNQ shows two weak bands at 2213 and 2224\,cm$^{-1}$ that
probably correspond to the modes visible in the F$_4$TCNQ spectrum
being red-shifted by 3-4\,cm$^{-1}$ due to the charge transfer. Based
on these numbers and on the $\nu$ values for fully ionized F$_4$TCNQ
\cite{F4TCNQ_meneghetti}, $\rho$ can be calculated with a
simple relationship given in Ref.~\cite{salmeronValverde1994}. The
results for the charge transfer degree are 0.12 in the case of
b$_{1u}$ and 0.09 for b$_{2u}$.
\subsection{Charge transfer transition energy}
\begin{figure}[htp]
\includegraphics[width=0.43\textwidth]{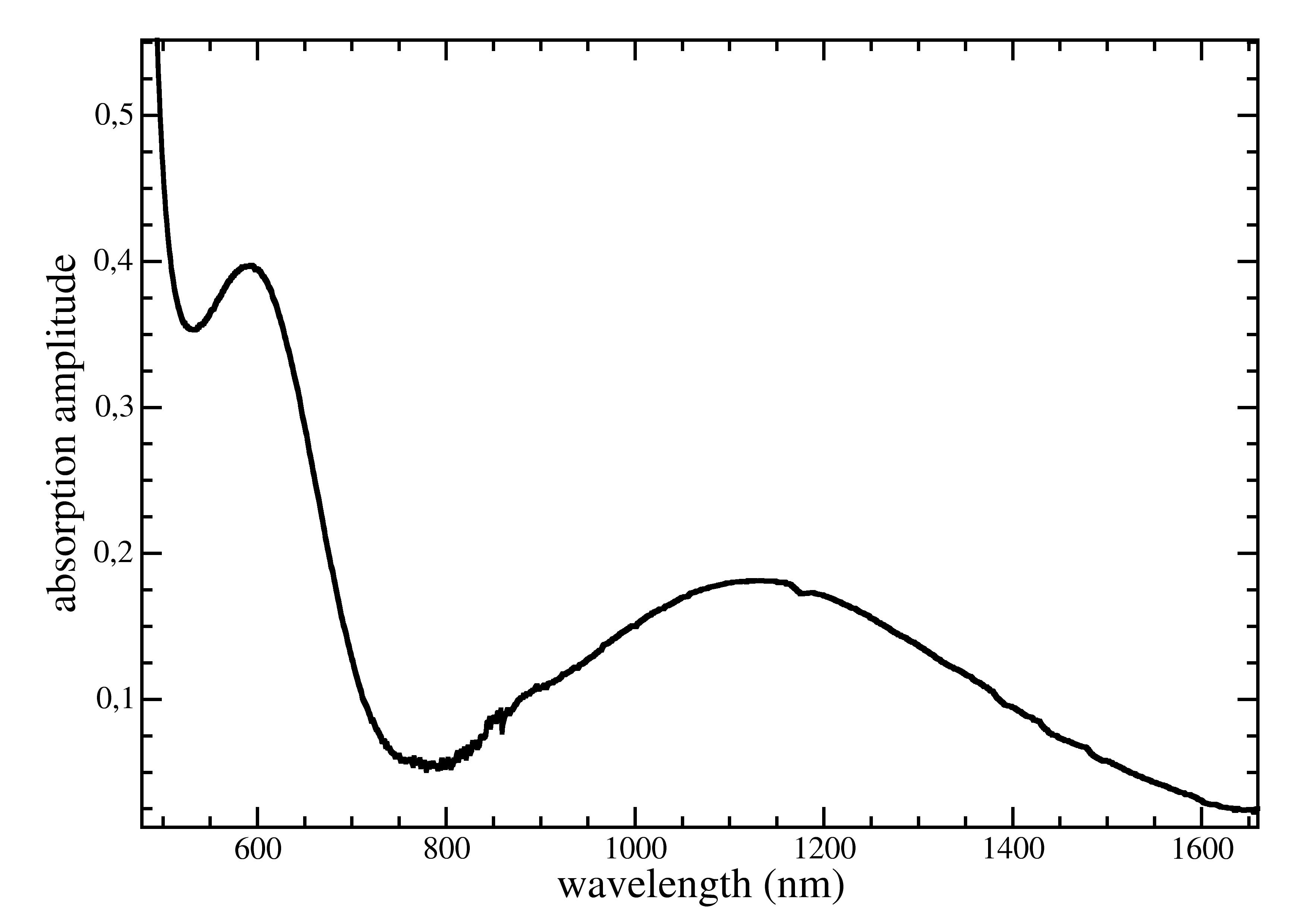}
\caption{Absorption spectrum of TMP-F$_4$TCNQ. Data were recorded from
  partly reacted TMP and F$_4$TCNQ in
  solution.}\label{fig:UVVIS_spektrum}
\end{figure}

Fig.~\ref{fig:UVVIS_spektrum} shows two absorption peaks in solution
in the region of interest: the left one at 592\,nm corresponds to an
energy of 2.1\,eV, the right one at 1132\,nm corresponds to
1.1\,eV. The latter presumably originates from the characteristic
charge transfer transition which is typically associated with a broad
peak shape like in our case.

The peak at 2.1\,eV could stem from intramolecular transitions in
donor or acceptor (neutral or ionized) molecules that generally appear
at higher energies than the CT transition. The same very likely
applies to the strong absorption partly visible in the leftmost
position of the figure. Here, one can refer to the measurements
performed on neutral TMP (thin films)\cite{TMPTCNQ_mainz} that showed
an optical gap of 3.167\,eV (392.7\,nm). Similar peak energies were
found for other donors/acceptors in the neutral or ionized
state. \cite{NI_discovery,optProp_TTFsalts,TMPTCNQ_mainz,CT_trans_chemMat}
\subsection{Electronic structure}
\begin{figure}[htb]
\includegraphics[width=0.45\textwidth]{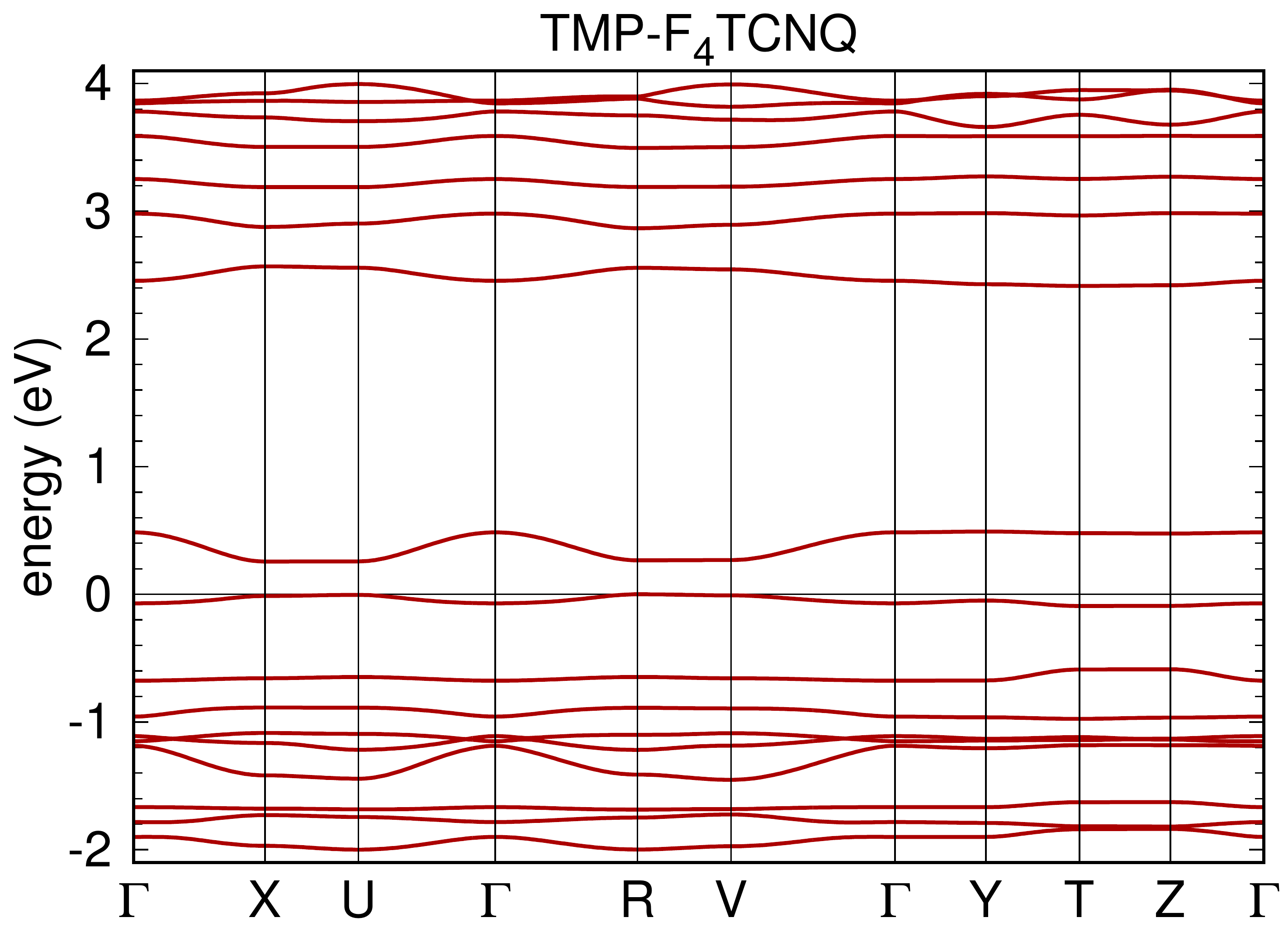}
\caption{Band structure of TMP-F$_4$TCNQ with a DFT gap of $E_g= 0.26$
  eV.}\label{fig:tmpf4tcnq_bs}
\end{figure}

\begin{figure}[htb]
\includegraphics[width=0.3\textwidth]{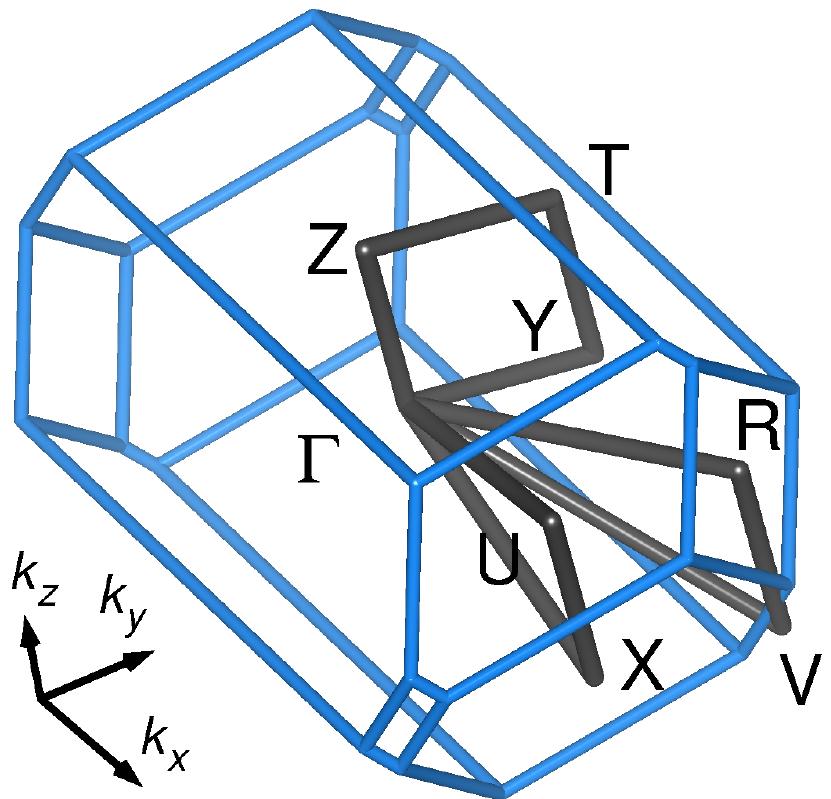}
\caption{Brillouin zone and chosen $k$ path for triclinic
  TMP-F$_4$TCNQ. }\label{fig:briZones}
\end{figure}

Fig.~\ref{fig:tmpf4tcnq_bs} shows the band structure of TMP-F$_4$TCNQ
in a large energy window. Coordinates of the high symmetry points are
$\Gamma=(0,0,0)$, $X=(0.5,0,0)$, $U=(0.5,0,0.5)$, $R=(0.5,0.5,0.5)$,
$V=(0.5,0.5,0)$, $Y=(0,0.5,0)$, $T=(0,0.5,0.5)$, $Z=(0,0,0.5)$ in
units of the reciprocal lattice vectors. The electronic structure for
TMP-F$_4$TCNQ is calculated in a primitive cell containing one formula
unit and it is characterized by well separated single bands of predominant TMP or
F$_4$TCNQ origin, respectively. These bands are a result of the hybridization between
the TMP highest occupied molecular orbital (HOMO) and the F$_4$TCNQ
lowest unoccupied molecular orbital (LUMO) (see Ref.~\cite{NEXAFS_mainz} for details).
 The band below the Fermi level  is
 mostly of TMP origin (donor) with some hybridization with F$_4$TCNQ
 while the band above the Fermi level
is of F$_4$TCNQ origin (acceptor) with some hybridization with TMP. The bands
are narrow and only show dispersion along paths where the $k_x$
coordinate is changing (for a visualization of the $k$ path, see
Fig.~\ref{fig:briZones}). This indicates that the system is very
one-dimensional along the {\bf a} direction which is the stacking
direction (compare Fig.~\ref{fig:structures}). The GGA band gap is
$E_g= 0.26$\,eV. Correlation effects beyond the GGA approximation,
like local electronic interactions $U$ or nonlocal Coulomb
correlations $V$, have not been included in the calculations and could
play a role in the estimation of the band gap. However, the DFT (GGA)
results should be reliable in providing the trend that TMP-F$_4$TCNQ
has a narrower gap than TMP-TCNQ (see below).

\begin{figure}[htb]
\includegraphics[width=0.45\textwidth]{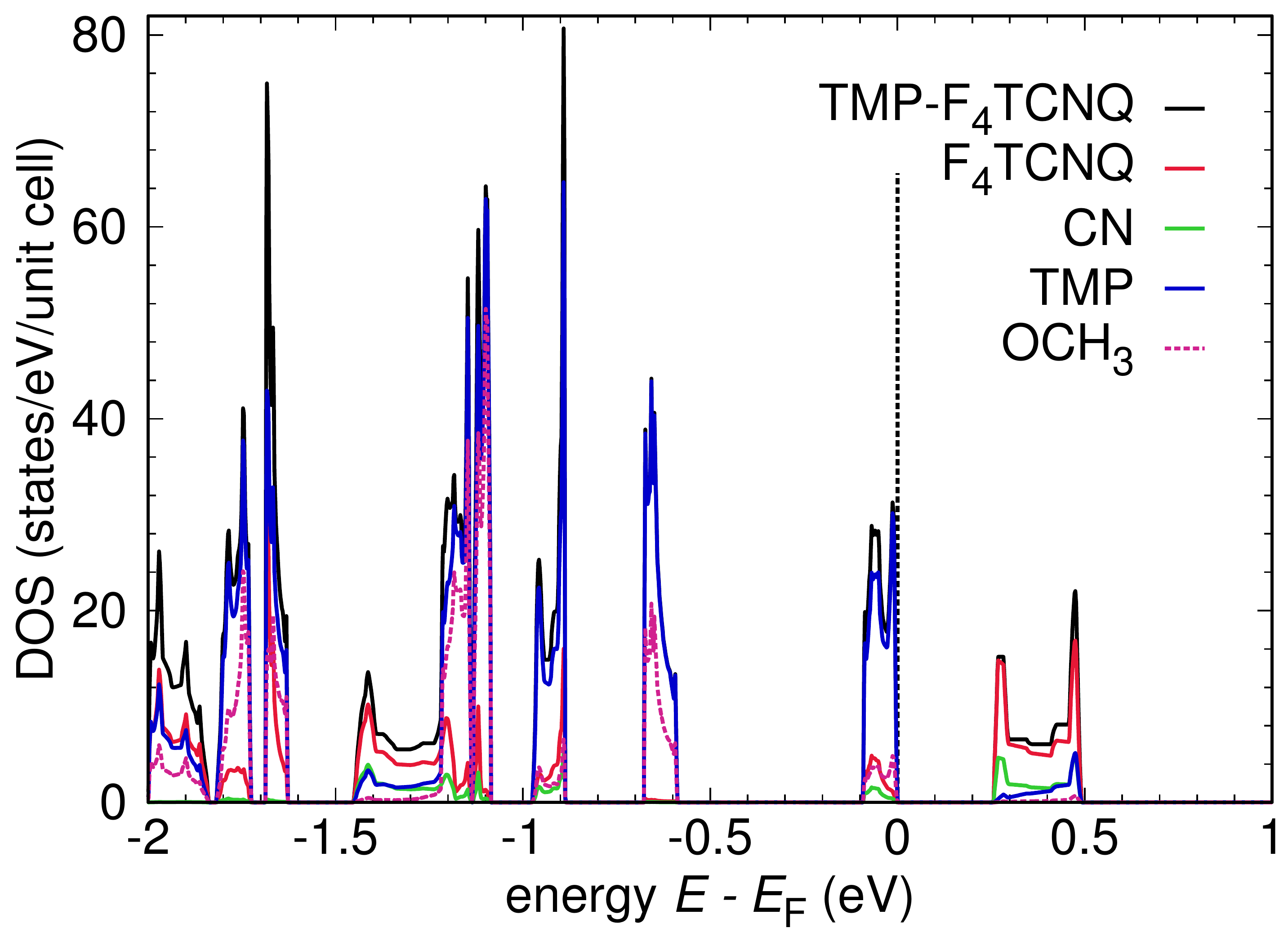}
\caption{Density of states of TMP-F$_4$TCNQ. The Fermi level is
  indicated by a dashed line. Contributions of TMP are shown in blue,
  of F$_4$TCNQ in red, of the cyano groups in green and of the methoxy
  groups in purple.  }\label{fig:tmpf4tcnqdos}
\end{figure}

In Fig.~\ref{fig:tmpf4tcnqdos} we show the density of states (DOS) for
TMP-F$_4$TCNQ. The effect of the charge transfer is seen in the lowest
unoccupied states  of F$_4$TCNQ which show some degree of hybridization
with the highest occupied states of TMP. Consequently, the highest
occupied states of TMP-F$_4$TCNQ have some admixture of
F$_4$TCNQ/cyano group character, and the lowest unoccupied states have
contributions from TMP.

The charge transfer can be quantified using the charges on the atoms
as obtained from the charge density; these numbers can be corroborated
by integrating parts of the density of states shown in
Fig.~\ref{fig:tmpf4tcnqdos}. Note that both ways to determine the
charge transfer involve different approximations and should not be
expected to give exactly the same result. Analysis of the charge
transfer yields a value of 0.20 electrons per formula unit from TMP to
F$_4$TCNQ. This agrees roughly with 0.24 electrons
F$_4$TCNQ-contribution to the highest occupied, TMP-derived bands of
TMP-F$_4$TCNQ (Fig.~\ref{fig:tmpf4tcnqdos}) and with 0.31 electrons
TMP-contribution to the lowest unoccupied, F$_4$TCNQ-derived
bands. Comparison of this result with the charge transfer obtained
from measurements (previous section) shows a disagreement that will be
discussed below.

\begin{figure}[htb]
\includegraphics[width=0.45\textwidth]{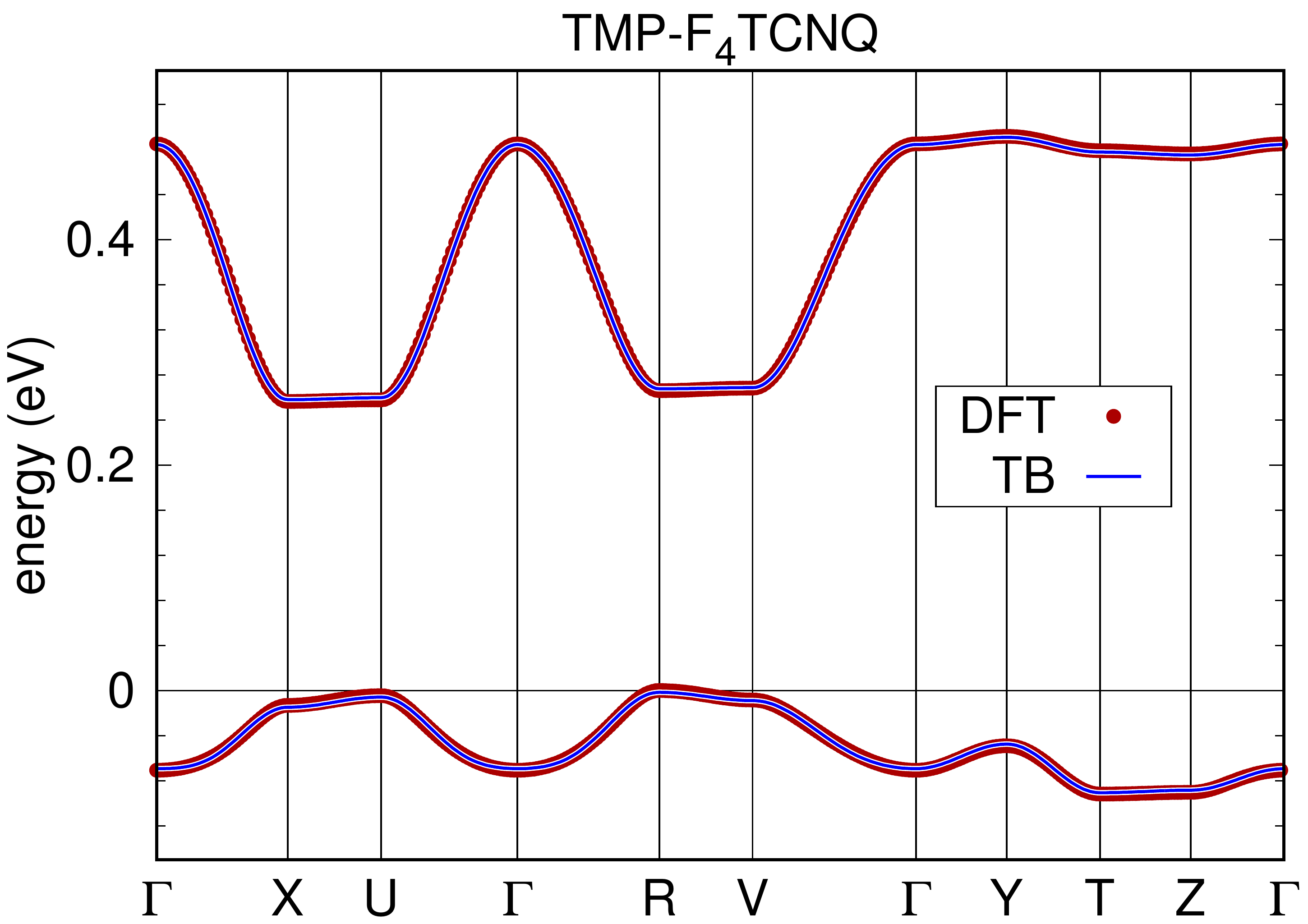}
\caption{Tight-binding bands for TMP-F$_4$TCNQ obtained from
  projective Wannier functions (solid line) shown together with the
  result of the DFT calculation.}\label{fig:tmpf4tcnq_tbbs}
\end{figure}

\begin{figure}[htb]
\includegraphics[width=0.45\textwidth]{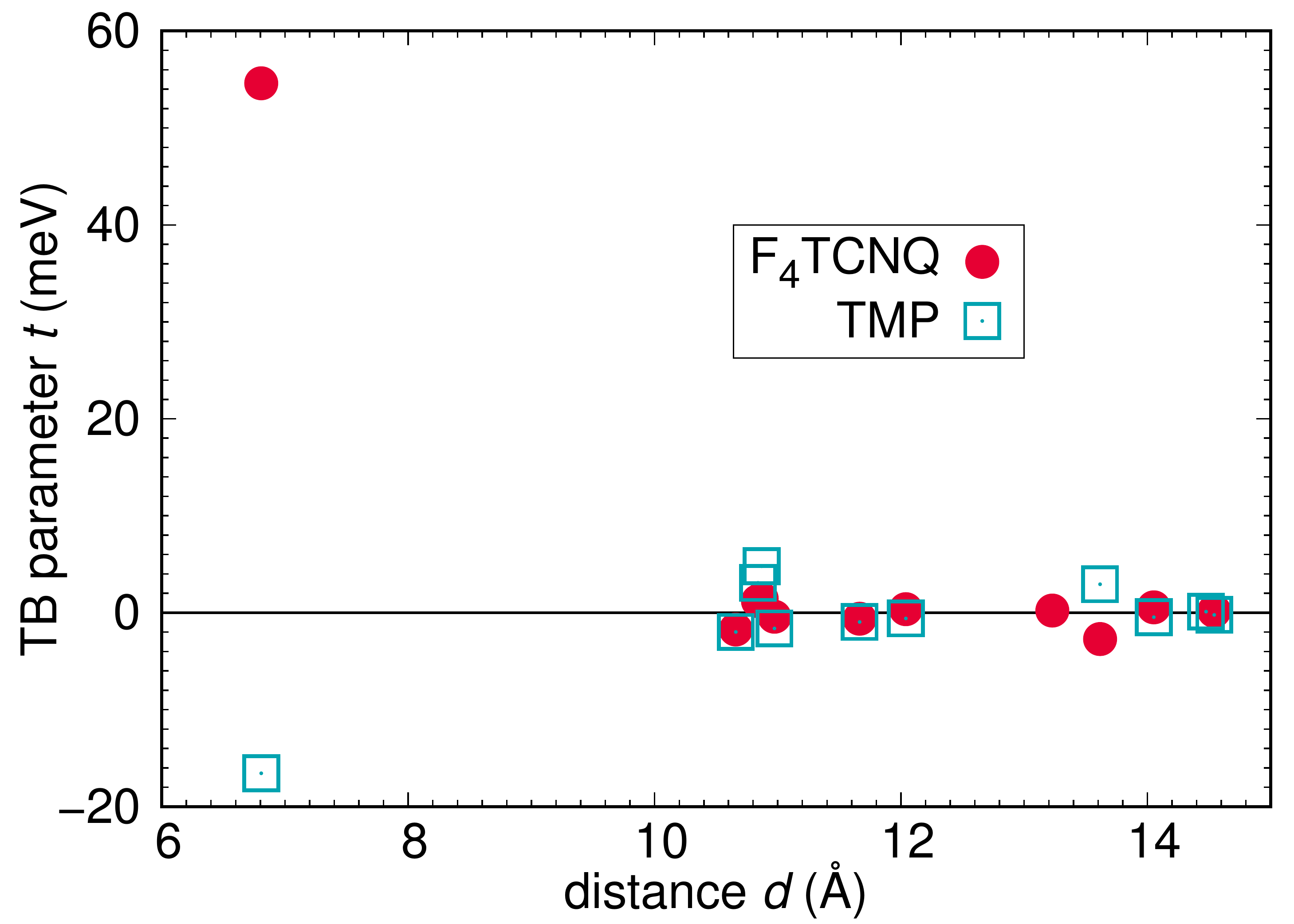}
\caption{Tight-binding parameters for TMP-F$_4$TCNQ. No hoppings
  between TMP-derived and F$_4$TCNQ-derived bands are
  found. }\label{fig:tmpf4tcnq_tbpar}
\end{figure}

We now determine the tight-binding Hamiltonian (compare
eq.~\eqref{eq:hamiltonian}) for the highest occupied and lowest
unoccupied bands of TMP-F$_4$TCNQ.  Fig.~\ref{fig:tmpf4tcnq_tbbs}
shows perfect agreement of the obtained tight-binding bands with the
DFT bands. The values of the tight-binding parameters obtained
from projective Wannier functions are plotted in
Fig.~\ref{fig:tmpf4tcnq_tbpar}. The largest contributions in both  valence
and conduction bands, correspond to
nearest neighbor hoppings between alike molecules along the
stacking {\bf a} direction, what is related to the strong
one-dimensional character of both bands. Note that there are no tight-binding
parameters connecting the two bands. This is a common feature in donor-acceptor systems.

\subsection{Uniaxial pressure simulation}
Eventually, it would be useful to develop methods to design charge
transfer salts, relying not only on the quantum chemical analysis of
binding energies for the donor and acceptor molecules, but taking into
account also the arrangement of the molecules in a crystal. We now
explore two aspects of the crystal structure that are important for
the charge transfer: The spacing of the donor and acceptor molecules
along the stacking axis and the respective arrangements of the charge distributions on the molecules. For that purpose, we fully relax within density functional theory with the GGA approximation the internal
coordinates of the TMP-TCNQ and TMP-F$_4$TCNQ structures for a small set
of reduced $a$ lattice parameters, thus simulating uniaxial pressure. Such an application of uniaxial
pressure or, alternatively, biaxial tensile strain, as often observed in thin
films\cite{Huth2014a}, could in principle be performed also in
experiment, but this is beyond the scope of the present investigations
and in this work we rather consider these effects as a computer
experiment that could allow us to predict, if application of pressure
may induce small chemical modifications of donor or acceptor molecules
that may be beneficial in the context of optimizing the charge
transfer degree.

\begin{figure}[!ht]
\includegraphics[width=0.45\textwidth]{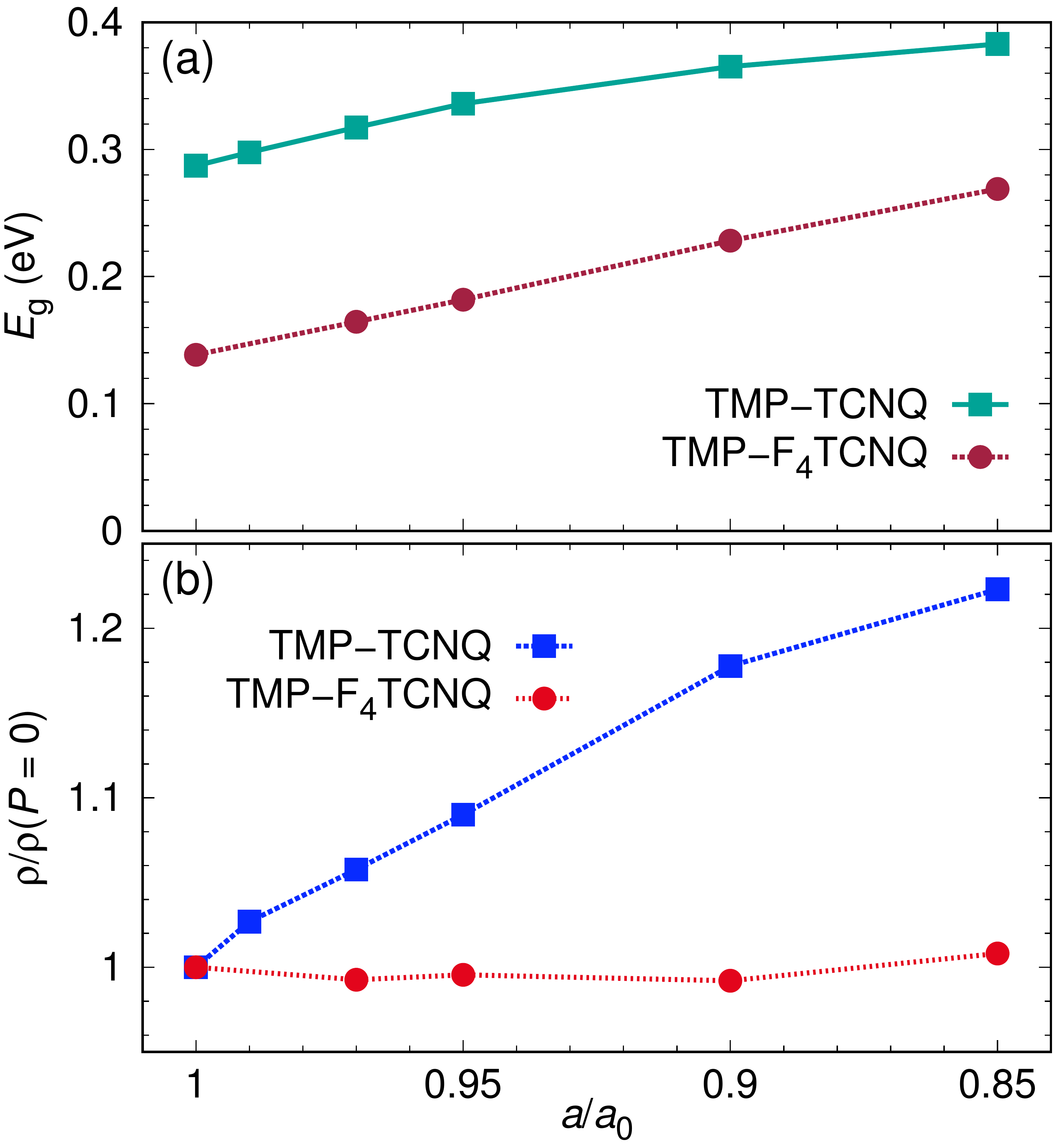}
\caption{Evolution of band gap (a) and charge transfer (b) from TMP to
  TCNQ (F$_4$TCNQ) as function of uniaxial compression along the
  stacking direction $a$. }\label{fig:ct}
\end{figure}

\begin{figure}[!ht]
\includegraphics[width=0.45\textwidth]{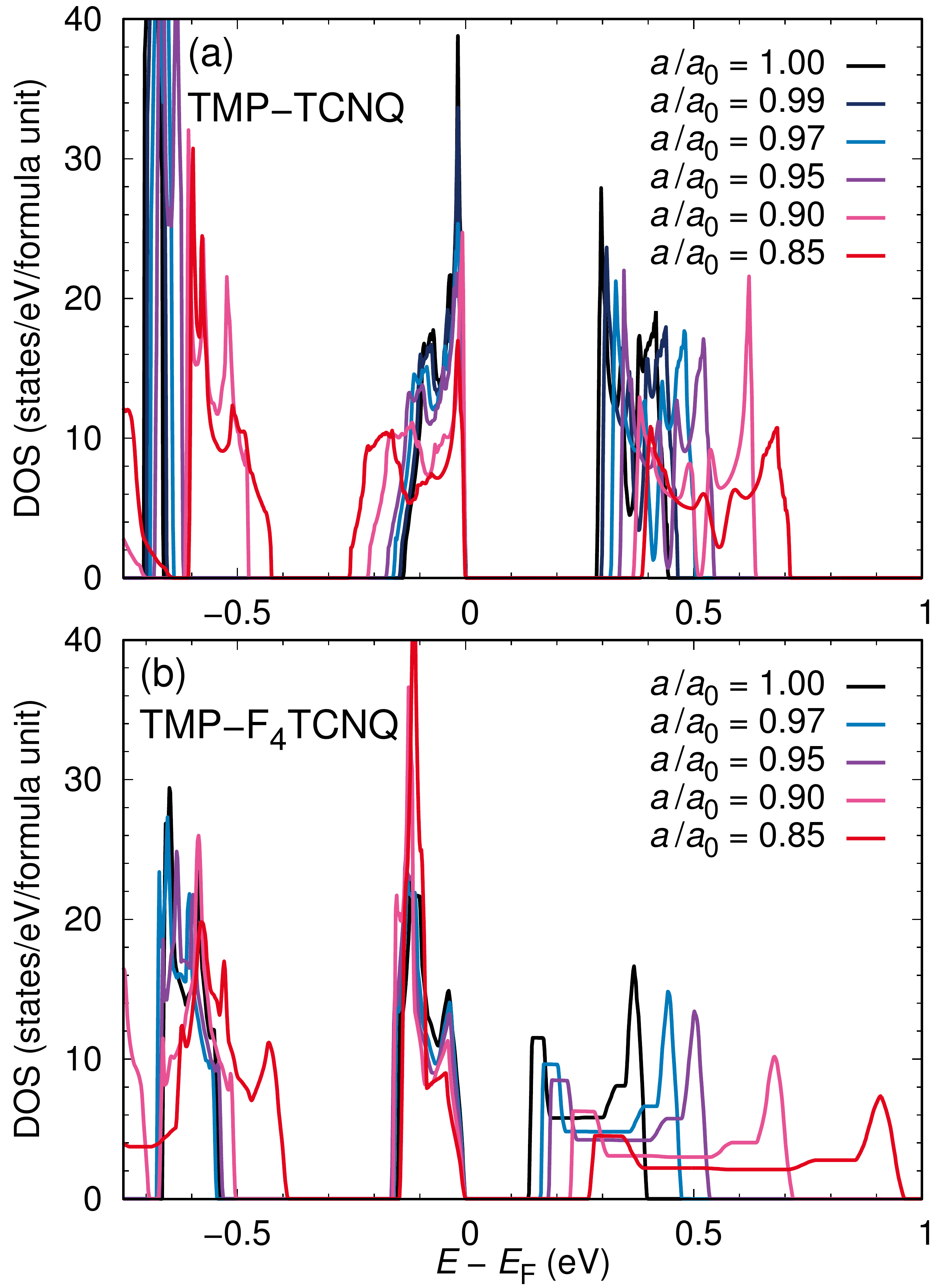}
\caption{Density of states (DOS) of (a) TMP-TCNQ and (b) TMP-F$_4$TCNQ
  for different degrees of uniaxial compression along the stacking
  direction $a$. }\label{fig:compress}
\end{figure}

We consider here the consequences of uniaxial strain in TMP-F$_4$TCNQ
and also in TMP-TCNQ for comparison. Fig.~\ref{fig:ct} shows the evolution of the
band gap, as well as the charge transfer from TMP to TCNQ in TMP-TCNQ
and from TMP to F$_4$TCNQ in TMP-F$_4$TCNQ as a function of reduced
$a$ lattice parameter; $a_0 = a(P=0)$ corresponds to the $a$ lattice
parameter at ambient pressure. Note that the structures for different 
$a/a_0$ where obtained by relaxing the internal coordinates of the complexes within DFT.
For TMP-TCNQ and $a/a_0=1$  the relaxed structure shows only small
differences with respect to the experimental structure. For the calculated
charge transfer in the relaxed structure we obtain an offset of about 0.2 with
respect to the value deduced from the DFT calculations relying on the experimental crystal structure. For TMP-F$_4$TCNQ the corresponding offset is about 0.1. In the representation of the uniaxial strain effects on the charge transfer, shown in Fig.~\ref{fig:ct}(b) the results are normalized to the value calculated for the relaxed structures at zero strain. Interestingly, we find different trends in TMP-TCNQ and TMP-F$_4$TCNQ: In TMP-TCNQ, the charge transfer
increases approximately linearly with compression along $a$, while in
TMP-F$_4$TCNQ it stays constant. For a more detailed analysis, we show
in Fig.~\ref{fig:compress} the densities of states for all values of
the $a/a_0$ ratio. As expected, overall band width increases as a
function of pressure in both cases. However, if we focus on the states
directly below the Fermi level that correspond mostly to the highest
occupied molecular orbitals of TMP, we see an important difference:
While the dispersion corresponding to these states significantly
increases with uniaxial compression in TMP-TCNQ, it remains nearly
unchanged in TMP-F$_4$TCNQ. In the next section we will discuss possible reasons for this different behavior related to the differences in the charge density arrangement of the two compounds.

\begin{figure}[!ht]
\includegraphics[width=0.45\textwidth]{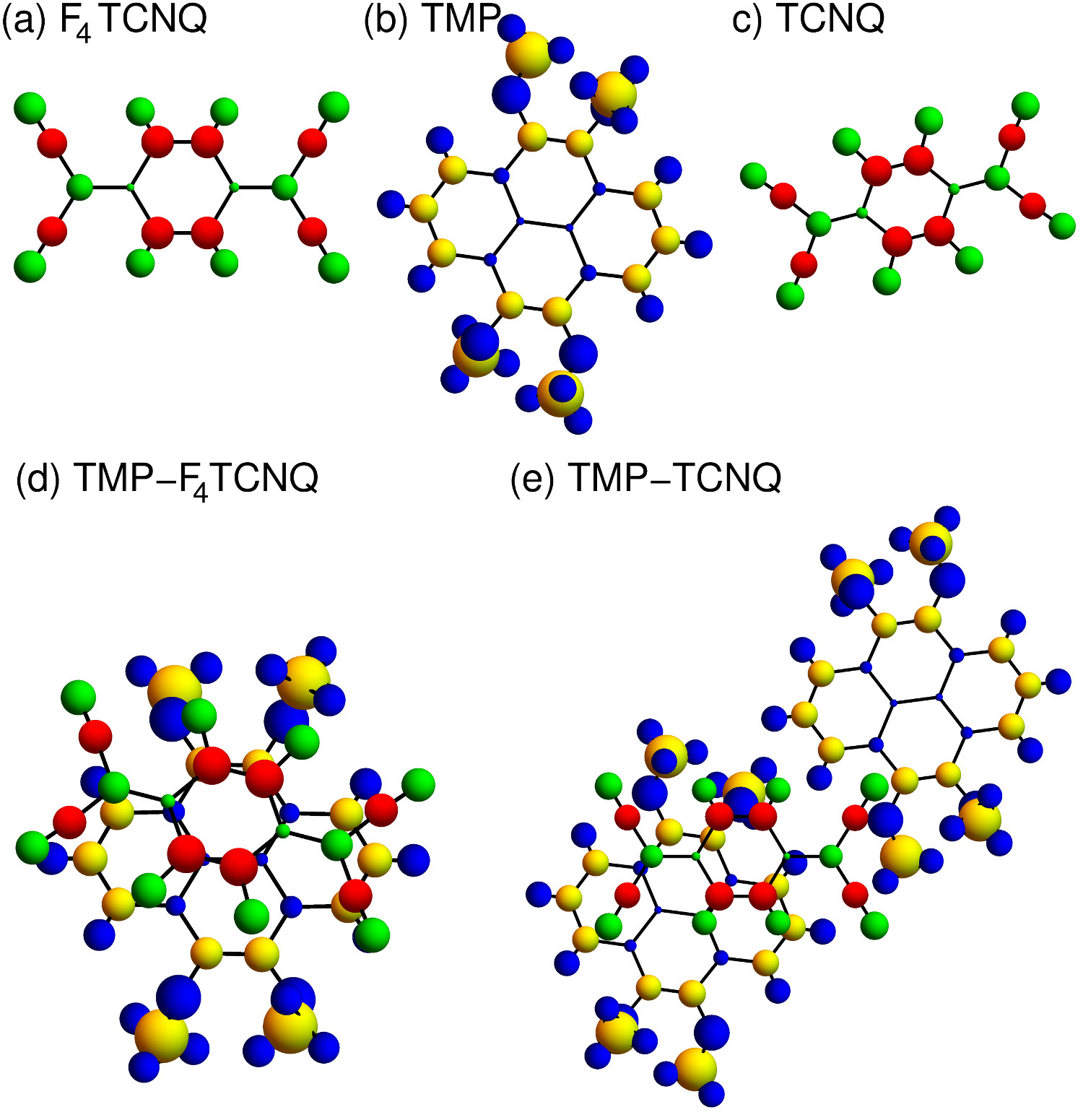}
\caption{Distribution of charges in (a) F$_4$TCNQ, (b) TMP, (c) TCNQ,
  (d) TMP-F$_4$TCNQ and (e) TMP-TCNQ.  Spheres are drawn with volume
  proportional to the excess charge (positive in blue, green and
  negative in yellow and red) on each atom. In (d) and (e), the actual
  alignment of donor and acceptor from the crystal structure is show.
}\label{fig:charges}
\end{figure}

\section{Discussion}
The crystal structure of TMP-F$_4$TCNQ with its simple and clear
stacking arrangement of donor-acceptor molecules gives rise to a
markedly reduced dimensionality of its electronic transport
characteristics. This is confirmed by both, our conductivity
measurements and band structure calculations. The dispersion visible
only along $k_x\parallel${\bf a} is reflected in an appreciable
increase of the conductivity along the stacking axis. This
one-dimensional character is a common phenomenon in CT systems but not
necessarily pronounced in mixed-stack CT. However, we observe that an
almost perfect one-dimensional geometry does not necessarily lead to
an optimized interaction that promotes maximum conductivity. The
semiconductor-like behavior of $\sigma$(T) below room temperature in
TMP-TCNQ is pronounced due to the rather high activation energies. The
conductivity is further limited by the rather small charge transfer,
if one takes into account that high $\sigma$-values should not be
expected for $\rho<$0.4.\cite{salmeronValverde1999}

In order to further analyze the influence of the geometry of the
stacking on the conductivity behavior as well as the role of the
specific acceptor F$_4$TCNQ in more detail, we compare in this section
the electronic properties of TMP-F$_4$TCNQ to
TMP-TCNQ~\cite{TMPTCNQ_mainz,NEXAFS_mainz}.

TMP-TCNQ shows a mixed-stack arrangement like TMP-F$_4$TCNQ
but the low dimensionality is far less pronounced~\cite{NEXAFS_mainz,Kawano2013a}.
 TMP and TCNQ
molecules are tilted against the stacking axis by slightly different
angles, i.e. their planes are less parallel compared to the
TMP-F$_4$TCNQ geometry. Within the stacks, they are shifted along {\bf
  b} and {\bf c} so that the geometrical centers of adjacent molecules
do not coincide when one plane is projected onto another. This shift becomes particularly apparent when comparing the distribution of excess charges viewed along the stacking axis, as is shown in Fig.~\ref{fig:charges}. By introducing the fluor in F$_4$TCNQ a singnificant increase of negative excess charge at the acceptor ring position occurs which seems to act as a strong docking point for the positive excess charge located at the ring close to the methoxy groups of the donor. This leads to a ring-centering effect in the donor-acceptor stacking of TMP-F$_4$TCNQ which has also the consequence of improving the planarity of the F$_4$TCNQ molecule. In TMP-TCNQ the planarity of the TCNQ structure is reduced when compared with
F$_4$TCNQ as the CN groups show stronger twists with respect to the
central ring and even the ring itself is not really planar. From these observations we draw two conclusions relating to the molecular overlap and the associated conductivty, as well as the expected behavior under applied uniaxial pressure along the stacking axis.

First, planarity is generally regarded as a relevant factor for higher
conductivity due to the increase of the molecules' overlap
integrals.\cite{organConduct} The reduced planarity and lateral offest in the ring stacking in TMP-TCNQ is therefor reflected by
significantly reduced conductivity values and the absence of an
observable anisotropy when compared to TMP-F$_4$TCNQ. In fact, the exact room-temperature value(s) of the
conductivity for different directions of a mm-sized TMP-TCNQ single crystal could not be
determined as its resistance exceeded the isolation resistance of our
setup. Nevertheless, we can state an upper limit of about 10$^{-11}$\,S/cm. The fundamental differences between TMP-TCNQ and TMP-F$_4$TCNQ are confirmed by the band structure calculations. TMP-TCNQ bands are narrower and show weak dispersion in all directions. The bandwidth is only 0.1 eV while it is 0.25 eV for TMP-F$_4$TCNQ.

Second, with regard to expected behavior of the charge transfer degree under uniaxial compression, the near-ideal planarity and ring-stacking of TMP-F$_4$TCNQ seems to preclude a significant increase of the donor-acceptor hybridization, as is clearly suggested by the results of the DFT calculations shown in Fig.~\ref{fig:ct}. For TMP-TCNQ, on the other hand, a significant increase of planarity of the acceptor molecules under uniaxial pressure may be expected, as well as possible lateral rearrangements leading to improved ring stacking. In this case, a clear tendency for a growing charge transfer degree should be expected, as is strongly suggested by the DFT calculations. We consider these predictions to be a robust feature despite of the fact that the calculated degrees of charge transfer, in particular for the fully relaxed structures, do not agree very well with the experimental values obtained from analyzing the CN vibration frequency shifts. We conclude this section with some comments in this direction.


Regarding the degree of charge transfer in TMP-TCNQ the CN vibration
showed a frequency shift from 2227 (pure TCNQ) to 2221\,cm$^{-1}$
indicating a charge transfer of 0.14. In this case, the
theoretical analysis of charges, relying on the experimental crystal structure, produced a nearly perfect agreement as
an overall charge transfer of 0.13 from TMP to TCNQ was
found. It agrees roughly with 0.16 electrons TCNQ contribution to the
highest occupied, TMP-derived bands of TMP-TCNQ, and with 0.21
electrons TMP contribution to the lowest unoccupied, TCNQ-derived
bands (experimental structure). In the case of TMP-F$_4$TCNQ the agreement between theoretical
(ca.\ 0.2-0.3) and the extracted values from infrared measurements (0.12)
is not as good. Considering that the absolute CN vibration frequency shifts are rather small, the error margins for the experimentally deduced charge transfer values are about 0.05. In addition, the observed changes in $\rho$ derived from the DFT calculations when allowing for full relaxation, in particular for TMP-TCNQ, indicate that already subtle changes in the crystal structure can lead to substantial changes. Taking this together, the deduced absolute values for the charge transfer degree in the TMP-TCNQ and TMP-F$_4$TCNQ systems have to be considered with caution. However, the predicted uniaxial pressure trends and their rationalization with a view to the stacking arrangement and excess charge distributions provide valuable insight into the somewhat counter-intuitive trends observed when going from TMP-TCNQ to TMP-F$_4$TCNQ. This work has shown that simple arguments that the significantly larger electron affinity of F$_4$TCNQ (ca. 5.2\,eV) in comparison to TCNQ (ca. 4.8\,eV) \cite{organConduct} should lead to an increased charge transfer do not hold. The present comparison also forms an interesting analogy to similar
investigations like the one by Torrance\cite{HMTTF_paper} who contrasted HMTTF-TCNQ with HMTTF-F$_4$TCNQ. In their study the different acceptors led to isostructural CT systems but with very different degrees of charge transfer. As a result HMTTF-F$_4$TCNQ becomes a Mott insulator while HMTTF-TCNQ is an organic metal.
\section{Conclusion}
We have investigated the structural and electronic transport
properties of the new charge transfer compound TMP-F$_4$TCNQ in the
form of single crystals. Both crystal and band structure show a
pronounced reduction of dimensionality as the nearly perfect
geometrical overlap of molecules along the mixed donor-acceptor stacks
leads to a one-dimensional transport characteristics. The predictions
are qualitatively confirmed by the measurement of anisotropic,
semiconductor-like electrical conductivity that is largest in stacking
direction. Based on IR spectroscopy a small charge transfer of about
0.1 was found while theoretical values as well as some
general considerations concerning the maximization of charge transfer
suggest a higher value. In comparison, the related CT complex
TMP-TCNQ, introduced in recent
publications\cite{TMPTCNQ_mainz,NEXAFS_mainz}, shows clear structural
differences. They result result in a more isotropic transport
behaviour with a significant overall reduction of electrical
conductivity owing to the loss of the apparently optimized molecular
overlap found in TMP-F$_4$TCNQ. At the same time our theoretical study on the evolution of the charge transfer degree
in TMP-(F$_4$)TCNQ suggests that for TMP-TCNQ a significant increase under unixal pressure along the stacking axis should occur whereas no such effect is expected for TMP-F$_4$TCNQ. Whether TMP-TCNQ may be a candidate for a pressure-induced neutral-ionic transition has to remain for future resolution.
\begin{acknowledgments}
  We thank V. Solovyeva, R. Rommel, M. Schmidt and S. Bek\"{o} for
  experimental support as well as helpful discussions. This project is
  funded through ``SFB/TR 49'' by the Deutsche Forschungsgemeinschaft.
\end{acknowledgments}
\bibliography{tmp_f4tcnq}
\end{document}